# Public Goods From Private Data - An Efficacy and Justification Paradox for Digital Contact Tracing

Andrew Buzzell
abuzzell@yorku.ca

Debate about the adoption of digital contact tracing (DCT) apps to control the spread of COVID-19 has focussed on risks to individual privacy (Sharma & Bashir 2020, Tang 2020). This emphasis reveals significant challenges to ethical deployment of DCT, but generates constraints which undermine justification to implement DCT. It would be a mistake to view this result solely as the successful operation of ethical foresight analysis (Floridi & Strait 2020), preventing deployment of potentially harmful technology. Privacy-centric analysis treats data as private property, frames the relationship between individuals and governments as adversarial, entrenches technology platforms as gatekeepers, and supports a conception of emergency public health authority as limited by individual consent and considerable corporate influence that is in some tension with the more communitarian values that typically inform public health ethics. To overcome the barriers to ethical and effective DCT, and develop infrastructure and policy that supports the realization of potential public benefits of digital technology, a public resource conception of aggregate data should be developed.

**Contact Tracing and COVID-19**

Since the successful development of contact tracing as a tool to help control diseases such as smallpox (Porco et al 2004) and tuberculosis (Begun et al 2013), public health authorities have had the power to compel individuals and organizations to provide data that can be used to analyze the movements and behaviour of an individual diagnosed with an infectious disease to identify possible incidences of transmission.

The virology of COVID-19 creates two kinds of scaling challenges that make manual contact tracing unfeasible. The mode of transmission is respiratory droplet spread, with some evidence of transmission via indirect surface contact, and the potential for aerosolized transmission in limited circumstances (Van Doremalen et al 2020). With a reproductive rate sufficient for exponential case growth, this creates a horizontal problem of resource scale, in the US alone it is estimated that over 100,000 full-time contact tracers (Watson et al 2020). The long period of infectivity, and in particular the period of asymptomatic transmission, creates a vertical scaling problem, where the amount of data required to conduct tracing for each individual is quite large, encompassing contacts with people and surfaces over a 14 day period.

DCT apps could mitigate the vertical problem by assisting recall, recording high fidelity data for each individual that can be retroactively queried to identify potential transmission, and the horizontal problem by automating much of the contact tracing process (Ferretti et al 2020). Even without a vaccine, effective DCT could allow public authorities to relax some of the severe restrictions that have been imposed, an important counterfactual when considering the justifiability of DCT programs (Mello & Wang 2020).

Most DCT proposals use Bluetooth Low Energy (BLE) radio networking technology present in smartphones, recording Received Signal Strength Indicator (RSSI) measurements to determine when devices are close together, and for how long. A database of device pairings and RSSI information is maintained on the device or a centralized server, and when one device is flagged as belonging to an infected individual, an algorithm can select from the database identifiers recorded while the individual may have been infectious, filter them by duration and signal strength, and produce a list of device ids that might be targeted for intervention of some type, such as testing or self-isolation.

As a sociotechnical system, DCT re-taxonomizes RSSI data as predictions of disease transmission risk and mandates actions, backed by public health authority. Justification for the ensuing actions depends in part on the reliability of the prediction. DCT faces serious efficacy challenges with both prediction and coverage, summarized in the supplementary material. When the non-causal proxies for transmission are too weakly correlated with actual transmission risks, or the individual or population coverage is insufficient or uneven, DCT can't perform the function of identifying infection risks effectively. While predictive problems might be mitigated by improving technology and aggregating additional data, coverage problems threaten the viability of DCT directly, and are least amenable to post-hoc correction. They require populations be persuaded to use the DCT app, and that hardware and software vendors cooperate with public health authorities to resolve barriers to adoption and usage, such as the need for software modifications to enable passive RSSI measurement.

**Efficacy as a Condition on Justification**

The exercise of coercive authority in the interests of public health is typically justified by the harm principle (Upshur 2002), that the action be necessary to prevent harm to others. It is further limited by the principle of least infringement (Childress et al 2002), that interventions which undermine privacy or autonomy must be the least burdensome alternative necessary to support the public health objective independently justified by the harm principle. Efficacy is therefore a necessary condition on justification, and any modulation of measures taken in response to other ethical concerns must maintain a level of efficacy consistent with claims that the intervention is a viable alternative (Allen & Selgelid 2017). For example, evidence that the pervasive use of face coverings in public significant reduces inter human transmission of COVID-19 (Zhang et al 2020) might justify the exercise of state power to make them compulsory, a significant limitation on autonomy, but one that is relatively low in costs and restrictions, compared to alternatives such as mass shelter-at-home orders. The efficacy of the less-restrictive alternative is high enough that the marginally better results from dramatically more severe restrictions are offset.

If responsiveness to ethical or legal requirements constrain implementation of DCT in ways that weaken its expected efficacy, this in turn undermines justification for coercive measures to encourage adoption. This might indicate a fundamental problem with the proposed intervention. Because DCT often triggers actions that further impact individual autonomy, such as quarantine, with the predictions it makes, efficacy is particularly critical. Moreover, because DCT has the potential to generate knowledge of risks that could save lives, decisions that dilute their epistemic capacity are themselves ethically salient (Dennett 1986).

**DCT and Privacy**

At a time of heightened public awareness of the privacy and security challenges presented by our digital data exhaust, DCT has been subject to intense scrutiny on privacy grounds. There is a growing awareness that our data can be used in contexts that we would not consent to, and which could harm our interests. We might agree to let an app to track our music listening habits to recommend playlists, but be dismayed to learn it can be used to make inferences about our mental health (Allen 2015, Greenberg et al 2016) - indeed digital phenotyping of aggregate data from wide range of sources can generate health data and de-anonymize individuals. (Martinez-Martin et al 2018, Sobhani 2019).

Even where we might grant consent to use our data in one context of analysis, interpretation and action, such as infectious disease control, we might not be able to foresee functions the data might be used for within it. Similar problems with informed consent arise in the context of genetic research (Lunshof et al 2008), where uncertainty about usage problematizes consent, a problem magnified under the socio-technical conditions in which digital data is collected and retained, which generates very little friction to such re-contextualization and re-taxonomization.

In light of these concerns it is not surprising that many DCT models have focused on privacy-by-design, with strict minimization of data collected and transmitted, strong anonymization, a prohibition on the use of additional data sources (such as GPS), and policies demanding regular deletion of data and restrictions on uploading data to central servers. Privacy-preserving DCT models have been extremely influential, as evidenced by the extent to which implementations have coalesced around privacy-preserving standards (Chan et al 2020, Li and Guo 2020, Tang 2020) such as MIT's Private Kit (MIT 2020), PEPP-PT (PEPP Team 20202) and DP-3T

(Troncoso et al 2020), and the extent to which technology platform providers and health institutions (World Health Organization 2020) have embraced this approach.

**The Exposure Notification API as a Theory of Public Health Authority Power**

Because the design of mobile operating systems prevents the passive collection of bluetooth data, the cooperation of vendors is necessary to build effective DCT apps. The dominant mobile operating system vendors, Apple and Google jointly and rapidly developed the "Exposure Notification API" (Apple & Google 2020) to support limited DCT capabilities. Access to the Exposure API is tightly controlled, and only one app can be deployed in a country. The vendors can disable and remove the app at any time. The app cannot use any data source except bluetooth RSSI data obtained via the Exposure API. The app cannot transmit this data to a central server. The Exposure API provides a methodology for the calculation of disease transmission risk which public health authorities configure by setting some pre-defined values.

The structure of the Exposure API expresses and enforces a policy perspective on the the relationship between public health authorities and citizens who use the products manufactured by Apple and Google. This treats data as private property, frames the relationship between individuals and governments as adversarial, entrenches technology platforms as gatekeepers and offers a conception of emergency public health authority as limited by individual consent and considerable corporate influence. This is an unconventional view - historically privacy is not signifiant constraint on manual contact tracing, and even strong legislation such as HIPPA recognizes the legitimate need for public health authorities access to protected health information (HIPPA 45 CFR 164.512)

Technology companies require a great deal of public trust to operate, as do governments and public health authorities. Because of the need for cooperation with governments to build DCT, vendors are exposed to highly publicized risks in the deployment of DCT, in terms of maintaining trust and also in avoiding additional regulation. The privacy preserving model serves vendor interests, allowing them to cooperate with public health authorities, thus avoiding regulatory or coercive measures, by limiting the possibility that the use of DCT apps breaks tacit or contractual agreements with their users that could damage already wavering public trust. (Newton 2020).

**Privacy-Preserving DCT Constrains Solutions to Efficacy Problems**

Critically, the Exposure Notification API prevents several actions that might be undertaken to improve the efficacy of DCT. Coverage problems that relate to contexts where smartphone ownership or physical possession is uneven could be partially remediated by aggregating other data, as could the predictive weaknesses of RSSI. The aggregation of data, including GPS, on central servers where it can be subject to further analysis and enrichment might also improve the epidemiological value of DCT (Mello & Wang 2020). Some countries have political, demographic and cultural characteristics that might favour the use of multiple apps, and data preservation may have future epidemiological value.

If privacy-maximizing constraints on DCT undermine efficacy, this in turn can weaken justification to deploy DCT at all. One might conclude that this is the correct outcome of ethical analysis of DCT, that it cannot be used ethically, because requirements needed to generate the efficacy required for public health objectives are unjustifiably invasive or coercive.

Alternately, one might wonder if this suggests that privacy-maximizing analysis is problematic. It is somewhat dismaying that a public health intervention that we have the technical means to deploy, which would be a much less restrictive alternative to measures currently in effect, becomes unjustifiable because of the restrictions necessary to ensure minimization of privacy risks.

Concerns about security and mission creep are only accidentally supportive of privacy maximization. While there are legitimate reasons to think that the socio-technical infrastructure DCT apps depend on are too insecure to trust, these are generally not inherent but are instead the results of implementation decisions, and in practice we are able to mitigate these to support many sensitive applications. There will be many examples of poorly implemented DCT, such as Qatar's which leaked personal data in QR-codes (Amnesty International UK 2020), but this does not mean that secure DCT is not possible.

One might also worry that governments will misuse the data down the road, but emergency public health legislation enacted in most jurisdiction has strict limitations that we should trust to function as intended. Even if we have upstream worries about the rule of law in some jurisdiction, this a distal problem, and not one that weighs in favour of the privacy-maximizing view generally.

**A Prediction and Coverage Paradox**

The problem of coverage efficacy is one of trust and influence as much as it is technical - adequate coverage and compliance depends in part on the public's willingness to cooperate. Discussion of DCT dominated by privacy and security concerns and messaging that prioritizes the protection of individual privacy influences public opinion, shaping the conditions of consent, which in turn affects the extent of coercive measures necessary to encourage adoption of DCT. Some jurisdictions plan to use choice architecture, such as defaulted opt-in, to encourage adoption and avoid coverage problems, but choice architecture, and the theory of libertarian paternalism that underlies it, depends on the absence of strong preferences (Sunstein 2015), but preferences can be shifted by the exercise of soft power in the information environment. Because the extent of the coercion is itself part of the justification calculus, changes in sentiment can impact justification.

An example of this relation between sentiment and justification is the effectiveness of anti-vaccination information operations (Johnson et al 2020, McKee & Middleton 2019, Wang et al 2019) which lead to a reduction in vaccine compliance in many jurisdictions, some of which have responded by increasing coercive regulation. This would be a difficult response to enact or justify if a majority of the population did not support it. One of the reasons why anti-vaccination propaganda, which is often produced and amplified by hostile entities, is particularly dangerous is that it can erode democratic mandate for the very actions that would mitigate the damage.

The paradox which arises for DCT is that increasing privacy protection in order to overcome constraints on justification undermines predictive efficacy to an extent that weakens justification to deploy DCT at all. But, to relax these protections to improve predictive efficacy conflicts with public sentiment (Milsom et al 2020), creating resistance to adoption that would exacerbate coverage efficacy problems, again weakening justification on efficacy grounds, but also increasing the justificatory burden because implementation against public sentiment raises the stakes in terms of autonomy impingement.

The remainder of this article explores a route to resolve this paradox by examining the conditions that make the privacy objections so difficult to overcome.

**Communitarian Bioethics, Principlist Technology Ethics**

Public sentiment against impingements on privacy necessary for DCT is grounded in legitimate fears of pervasive security problems with the socio-technical infrastructure. The litany of security and privacy problems with DCT applications that have already been deployed (Privacy International 2020) reinforce this.

However, this sentiment is also shaped by an increasingly prominent public discussion of technology ethics that is framed in a way that sits uneasily alongside the values that inform public health ethics. A dominant strain of technology ethics, as exemplified by legal expressions such as the EU's GDPR and California's CCPA and many AI ethics charters and codes of conduct (Jobin & Vayena 2019) resembles a format that, in bioethics, came to be known as "principlism" (Beauchamp & Childress 2001, Clouser & Gert 1990). This is the view that minimal set of principals, usually autonomy, non-maleficence, beneficence, and justice, supply the analytical machinery needed to approach ethical problems. It is criticized on the grounds that it does not specify an ordering, which instead is often inherited from the context of application, which tends to privilege the liberal individualist emphasis on autonomy, and which is unable to fully articulate principles such as beneficence beyond self-interest. (Callahan 2003). Applied technology ethics has a tendency to generate trade-off dilemmas, such as that between innovation and precaution, or privacy and public goods, because, as with principlism in bioethics, it does not supply a decision procedure for conflict resolution. This is particularly challenging when institutions that produce technological artefacts and systems struggle with "...onboarding external ethical perspectives..." (Metcalf & Moss 2019) that conflict with tacit and explicit internal norms. Our underlying moral interest in applied ethics demands more than compromise and consilience, rather, as Callahan puts it "[s]erious ethics, the kind that causes trouble to comfortable lives, wants to know what counts as a good choice and what counts as a bad choice" (Callahan 2003).

The "communitarian turn" in bioethics arose in part because capabilities emerging in genetic research created opportunities for public goods that could only be ethically realized once focus on individual interests yielded to more communitarian principles such as solidarity and public benefit. (Chadwick 2011). Predictive genetic analysis that might benefit an individual, their family, and their community, now and in the future, exposes information that might be prejudicial to the individual's interests, for example, by interfering with their ability to acquire health insurance (Fulda & Lykens 2006, Launis 2003) The extended value of genetic data over long timelines and across unforeseeable applications problematizes the coherence and applicability of autonomy protections such as informed consent. An ethical framework that could motivate policy and regulation to enables the pursuit of these opportunities for public good required the integration of communitarian values.

Likewise, public health ethics introduces principles such as solidarity, proportionality, and reciprocity alongside the four core principles of biomedical ethics (Coughlin 2014, Lee 2012, Schröder-Bäck et al 2014), communitarian values that reflect the fundamentally shared object of concern, and further expose the limits of analysis that privileges autonomy. Communitarian and distributive considerations can help resolve some of the ordering problems principlist technology ethics inherits from the liberal individualist context it operates within, helping to resolve tradeoffs by giving greater weight to shared values and common goods.

**A Public Resource Approach**

If DCT cannot be deployed in a way that is ethical and effective, this is an unfortunate loss of a significant public health opportunity. The barriers to remediation run deeper than privacy-preserving technical measures, and stem from the need to develop a conception of aggregate personal data as a public resource.

The Exposure Notification API encodes and enforces a privacy and autonomy maximizing model of DCT, essentially privatizing a public health policy concern. One justification for this is that corporations are enabling their users to protect their personal property, or adhering to a contractual obligation (Taddeo & Floridi, 2016). Traditional contact tracing treats our personal data as a potential public resource, with synchronous consent and access procedures triggered by the identification of transmission risk, whereas DCT treats it as a de-facto public resource with aways-on consent and access. DCT provides public benefits based on data collected from many individuals who might never have an elevated risk. Its value is at the population level, and we would accept impingement on our privacy for the good of the community. Although privacy is usually regarded as a paradigmatically individual concern, communitarian approaches to privacy (O'Hara 2010, Floridi 2017) argue that groups can have privacy rights, and that privacy is fundamentally a common good, where its value and limits are in reciprocal tension with other community values.

Technology companies profit from the value they extract from aggregate data, which depends on pervasive access to individual data in ways that frequently compromise privacy. Aggregate data is exponentially more economically and informationally valuable than that of the data of any one individual, and confers signifiant soft power to influence public sentiment, and hard power to control access to data and generate economic opportunities. But it is not clear that the equivocation between personal data as the private property of an individual, and aggregate data as the private property of the collector, is justified. Napoli (2019) argues that "...whatever the exact nature of one's individual property rights in one's user data may be, when these data are aggregated across millions of users, their fundamental character changes in such a way that they are best conceptualized as a public resource" (Napoli 2019). If aggregate data is substantially and uniquely distinctive, this supports the application of public trust doctrine, which is based on the idea that "...because of their unique characteristics, certain natural resources and systems are held in trust by the sovereign on behalf of the citizens" (Calabrese, 2001), such a the public broadcast spectrum.

The exploration of a communitarian approach to applied technology ethics and the articulation and assertion of a public resource rationale applicable to the data we generate by engaging with digital technology and services could enable policy and regulation that would directly address the barriers to effective and ethical DCT. This could expand regulatory and policy measures to ensure the safe handling of sensitive data, foster the enabling conditions for the realization of opportunities to use aggregate data for public good, and help reverse the centralization of decisive power over public policy in the hands of multinational technology corporations. Where policy and legislation such as the GDPR, especially through the DPIA process, identifies and protects risks to individual interests, methodologies to identify and protect opportunities in the public interest lag behind, as the barriers to DCT implementation illustrate.

**Supplementary Material - DCT's Efficacy Challenges**

**Inherent efficacy challenges**: The virology of COVID-19, so far as it is understood, makes this re-taxonomization problematic, because the mode of transmission and infectivity is such that there is only a weak likelihood that any particular contact detected by DCT results in transmission, whereas for disease such as tuberculosis or HIV/AIDS, it is easier to identify exposure events with high transmission probability. The contact/transmission link is also problematic due to the potential for transmission via indirect surface contact.

**Reliance on smartphones**: There are socioeconomic confounds related to smartphone ownership and use that will skew representation and the ability to install and update DCT apps. Life patterns in some populations generate periods of contact with others when smartphone are not present, and some forms of employment generate a large number of contacts with others, which may or may not actually correspond to increased risks of transmission. Evidence for non-nosocomial transmission in Japan shows primary cases in several contexts where smartphones are frequently not on our persons or turned off, such as music events and gyms (Furuse et al 2020).

**Bluetooth RSSI as a proxy for exposure**: There are efficacy problems with the core technology. RSSI measurements map only weakly to transmission risk, because BLE radio signals travel through walls and barriers used in public spaces to specifically to prevent droplet spread. RSSI is stronger when we walk side-by-side than following one another, is weakened when phones are in pockets while sitting around a table, and is sensitive to many idiosyncratic features of indoor environment (Leith & Farrell 2020). There are also considerable differences in RSSI measurement for different devices and different mobile operating systems (BlueTrace 2020), which introduces socio-economic confounds.

**Security:** Efficacy can be further undermined by deliberate exploitation of security vulnerabilities (Vaudenay 2020) and even simple circumvention such as the display of screen captures instead of running apps, as has been observed in India with mandatory Aarogya Setu app (Clarence 2020) undermines the public health value of DCT.

**Individual and Population Coverage:** At the population level, DCT apps would have to be in use by 60% of a population (Servick 2020) to be effective, a challenge that lead Singapore to consider making their app mandatory, a proposal later abandoned due to implementation challenges (Mahmud 2020). Various jurisdictions have considering opt-in, out-out, and incentivization schemes to encourage uptake.

At the individual level, coverage involves the extent of the individual's activities and behaviours that are accurately captured by the DCT app. Aside from issues related to smartphone ownership and presence described above, mobile phone operating systems place limits on the ways apps can access bluetooth radios, often requiring apps be open and in use - even an individual who has installed the app and has their phone at all times would produce little useful DCT data in this case. This problem is in fact a critical barrier to effective DCT, and requires the cooperation of operating system vendors to remediate, and requiring users to keep their phones open and the apps on-screen is not viable.